\documentclass[aps,prer,twocolumn,showpacs]{revtex4-1}
\usepackage{amsfonts}
\usepackage{amsmath}
\usepackage{amsthm}
\usepackage{graphics}
\usepackage{graphicx}
\usepackage{subfigure}
\usepackage{amssymb}
\usepackage{graphics}
\usepackage{graphicx}
\usepackage{epstopdf}
\usepackage{color}
\usepackage{subfigure}
\usepackage{url}
\usepackage{xfrac}

\def\x{{\bf x}}

\def\k{\textbf{k}}

\def\v{\textbf{v}}
\def\kmax{k_{\rm max}}
\newcommand\bet{{g}}
 
\newcommand\alps{{\frac{\hbar^2}{2m}}}

\newcommand\dertt[1]{ \frac{\partial{ #1}}{\partial t} }
\newcommand\gd{\mbox{${\bf \nabla}^{2}$}}

\begin{document}

\title{Grid superfluid turbulence and intermittency at very low temperature}

\author{Giorgio Krstulovic$^1$}
\affiliation{
Laboratoire J.L. Lagrange, UMR7293, Universit\'e de la C\^ote d'Azur, CNRS, Observatoire de la Côte d'Azur, B.P. 4229, 06304 Nice Cedex 4, France}


\begin{abstract}
Low-temperature grid generated turbulence is investigated by using numerical simulations of the Gross-Pitaevskii equation. The statistics of regularized velocity increments are studied. Increments of the incompressible velocity are found to be skewed for turbulent states. Results are later confronted with the (quasi) homogeneous and isotropic Taylor-Green flow, revealing the universality of the statistics. For this flow, the statistics are found to be intermittent and a Kolmogorov constant close to the one of classical fluid is found for the second order structure function.
\end{abstract}
\maketitle

\section{Introduction}

Superfluid turbulence has been largely studied in the last decades, especially thanks to the progress achieved in experimental technics. Today it is possible to create turbulent Bose-Einstein condensates (BEC)\cite{Henn2009}, visualize and track quantum vortices in BECs \cite{Serafini2015_VortexBEC} and $^4$He \cite{Bewley2006,MantiaVisPart,la2013lagrangian}, and study Lagrangian dynamics by using tracers \cite{Gao2015,Zmeev2013} in $^4$He. As in classical $3D$ hydrodynamic turbulence  \cite{Frisch1995}, a turbulent Kolmogorov cascade is observed at scales in between the energy injection scale and mean inter-vortex distance \cite{maurer1998local,salort2010turbulent}. 
 At scales smaller than the inter-vortex distance, the quantized vortices can not be considered as a continuous field and different mechanisms appear to be relevant to carry the energy to scales small enough to be dissipated by phonon emission \cite{Vinen2002,Walmsley2007}. Numerical simulations of different models confirm this scenario \cite{Nore1997a,Yepez2009,Salort2012,Baggaley2011a}. In classical and superfluid three-dimensional turbulence, energy is usually injected at large scales by different types of forcing. In classical hydrodynamic turbulence, one of the most standard ways is by a fluid flowing through a grid \cite{comte1966use,pope2000turbulent,stalp1999decay}. When increasing the mean flow, the fluid behind the grid develops a series of instabilities creating a turbulent wake. Such classical experiments have been also performed during the last decade using superfluids as  $^4$He \cite{Zmeev2015,Bradley2012}  and  $^3$He \cite{Bradley2005}. 

In this {work} we investigate low-temperature superfluid turbulence generated by a moving grid using the Gross-Pitavskii equation (GPE).  Statistics of (regularized) velocity increments are analyzed. The results are later confronted with homogeneous and (quasi) isotropic turbulent flow generated by the so-called Taylor-Green flow \cite{Nore1997}. Statistics of velocity increments are shown to be universal. An estimation of the dissipation energy rate leads to a measurement of the Kolmogorov constant close to the one observed in classical turbulence. Finally, the increments of the incompressible velocity are found to be intermittent.

The GPE describing a homogeneous BEC of volume $V$ with (complex) wave-function $\psi$ is given by
\begin{equation}
i\hbar\dertt{\psi} =- \alps \gd \psi + \bet|\psi|^2\psi ,
\label{Eq:GPE}
\end{equation}
where $m$ is the mass of the condensed particles and $g=4 \pi a \hbar^2 / m$, with $a$ the $s$-wave scattering length. Madelung's transformation $\psi({\bf x},t)=\sqrt{\frac{\rho({\bf x},t)}{m}}\exp{[i \frac{m}{\hbar}\phi({\bf x},t)]}$ relates the wave-function $\psi$ to a superfluid of density $\rho({\bf x},t)$ and velocity ${\bf v}={\bf \nabla} \phi$, {where $\phi$ is the phase of the wave-function}.  $\kappa=h/m$ is the Onsager-Feynman quantum of velocity circulation around the $\psi=0$ vortex lines. When Eq.\eqref{Eq:GPE} is linearized around a constant $\psi= \hat{\psi}_{\bf 0}$, the sound velocity is given by $c={(g| \hat{\psi}_{\bf 0}|^2/m)}^{1/2}$ with dispersive effects taking place at length scales smaller than the coherence length $\xi={(\hbar^2/2m|\hat{\psi}_{\bf 0}|^2g) }^{1/2}$, which also corresponds to the vortex core size ~\cite{Footenote1}. Using the Madelung transformation {(see \cite{Nore1997a} for details)} the energy {term (per unit of volume)} {$E=(\hbar^2/2mV)\int|\nabla \psi|^2 d\x$} can be rewritten as {$E=E^{\rm I}+E^{\rm C}+E^{\rm Q}=(1/2V)\int (|{\bf v}^{\rm I}|^2+|{\bf v}^{\rm C}|^2+|{\bf v}^{\rm Q}|^2)d\,\x$}, where
\begin{equation}
{\bf v}^{\rm I}=\mathcal{P}_{\rm inc}[\sqrt{\rho}\nabla \phi],\hspace{.1cm}{\bf v}^{\rm C}=\sqrt{\rho}\nabla \phi-{\bf v}^{\rm I},\hspace{.1cm}{\bf v}^{\rm Q}=\frac{\hbar}{m}\nabla{\sqrt{\rho}},\label{Eq:Velos}
\end{equation}
with $\mathcal{P}_{\rm inc}[\,\cdot\,]$ the projector onto the space of divergence-free fields. The super-index stand for incompressible (I), compressible (C) and quantum (Q) velocities. These fields are all regular at {the} vortex position as {they} are regularized by the term $\sqrt{\rho}$ \cite{Nore1997}. The velocity ${\bf v}^{\rm I}$ contains the contribution of vortices, whereas ${\bf v}^{\rm C}$ and ${\bf v}^{\rm Q}$ are related to waves, since they are by construction potential flows. The energy spectra are defined as {$E^{ \Lambda}_k(z)=\frac{1}{2}\int_{{|\bf p|}=k} |\widehat{{\bf v}^{ \Lambda}}({\bf p},z)|^2\mathrm{d}^2{\bf p}$, where $\widehat{{\bf v}^{ \Lambda}}$ is the Fourier transform of ${\bf v}^{\rm  \Lambda}$ in the plane perpendicular to the mean flow at a given distance $z$ from the grid. The superscript $ \Lambda$ stands for I,C and Q.}

In the simulations presented in this {work}, the mean density $m N/V$ is fixed to $1$ and the physical constants in Eq.\eqref{Eq:GPE} are determined by the values of $\xi$ and $c=1$. The quantum of circulation is given by $4\pi c\,\xi/\sqrt{2}$. Numerical integration of Eq.\eqref{Eq:GPE} is performed by using a fully de-aliased pseudo-spectral code. The domain is periodic in all directions. The perpendicular (respect to the direction of the mean flow) size of the domain is denoted by $L_\perp$ and the parallel one by $L_\parallel$. The grid is modeled by a strong repulsive potential $V_{\rm Grid}(\x)$.  The grid is characterized by the diameter of the rods  $a$ and the distance between the rods $D$.  {A sketch of the grid is shown in Appendix \ref{Appendix:Model}.} The fluid is initially at rest. The system is then advected with a velocity $v_0$ that is slowly increased from zero up to its final value. During this process, local dissipation is included far from the grid to reduce the sound emitted during the transient (see {Appendix \ref{Appendix:Numerics} for more details on numerics and methods}). Different grids, Mach numbers $M=v_0/c$ and resolutions are studied (see Table \ref{Table:RUNS}).
 \begin{table}[h]
\begin{tabular}{| c || c  |  c  | c | c |  c |c| c || c  |  c  | c | c |  c |}
  \hline
  \hline
Run 	& $L_\perp$	& $L_\parallel$	&$\frac{L_\perp}{D}$	& $N_\perp$	& $N_\parallel$ &&Run 	& $L_\perp$	& $L_\parallel$	&$\frac{L_\perp}{D}$	& $N_\perp$	& $N_\parallel$				 \\ \hline
a1    	&  $170$ 		& $683$		&	$9$			& $128$ 		& $512$		&&b1    	&  $341$	 	& $683$		&	$17$			& $256$ 		& $512$					 \\
a2   	&  $170$ 		& $683$		&	$5$			& $128$ 		& $512$		&&b2    	&  $341$ 		& $683$		&	$9$			& $256$ 		& $512$					 \\
a3    	&  $170$ 		& $683$		&	$7$			& $128$ 		& $512$		&&b3    	&  $341$ 		& $683$		&	$7$			& $256$ 		& $512$					 \\
a4    	&  $170$ 		& $683$		&	$3$			& $128$ 		& $512$		&&b4    	&  $341$ 		& $683$		&	$5$			& $256$ 		& $512$					 \\
\hline
c1    	&  $683$ 		& $683$		&	$7$			& $512$ 		& $512$		&&tg   	  &  $512$		& $512$		&	-			& $512$		& $512$					 \\
 \hline
 \end{tabular}
 \caption{List of runs. $L_\perp$ and $L_\parallel$ are the sizes of the domain and $N_\perp$ and $N_\parallel$ the corresponding resolutions. $D$ is the distance between the rods (see Appendix \ref{Appendix:Model} for details). The diameter of the rods is $a=2\xi$ for all runs. Lengths are expressed in units of the healing length $\xi$. The Mach number is $M=u_0/c=0.8$ for all runs except for a1m and a2m with $M=0.6$. For all grid runs $\xi=.75L_\perp/N_\perp$. For the Taylor-Green run (tg) $L_\perp=L_\parallel=L$, $N=N_\perp=N_\parallel$ and $\xi=L/N$. No symmetries are enforced.
 \label{Table:RUNS}}
  \end{table}
Note that such a periodic configuration mimics recent grid turbulence experiments with $^4$He in a ring \cite{Zmeev2015}, and similar ideas have been used in 2D to study the possibility of an inverse cascade \cite{Reeves2013}. We also study (quasi) homogenous and isotropic turbulence generated by the Taylor-Green flow {(see Fig.\ref{Fig:3DSnapshotsTG} below)}. This standard flow consists of a number of vortex rings and it develops a turbulent tangle followed by a small-scale thermalization of sound waves (due to the Galerkin projection of GP equation). Vortices exchange momentum and energy with the thermalized waves mimicking mutual-friction effects. These generic properties of the GP model {are} also expected to be present in the grid simulations. Note that the thermalization and its associated effects also occur independently of the spectral cut-off if dispersive effects are important \cite{Krstulovic2011a,Shukla2013}.

\begin{figure*}
\includegraphics[width=0.99\textwidth]{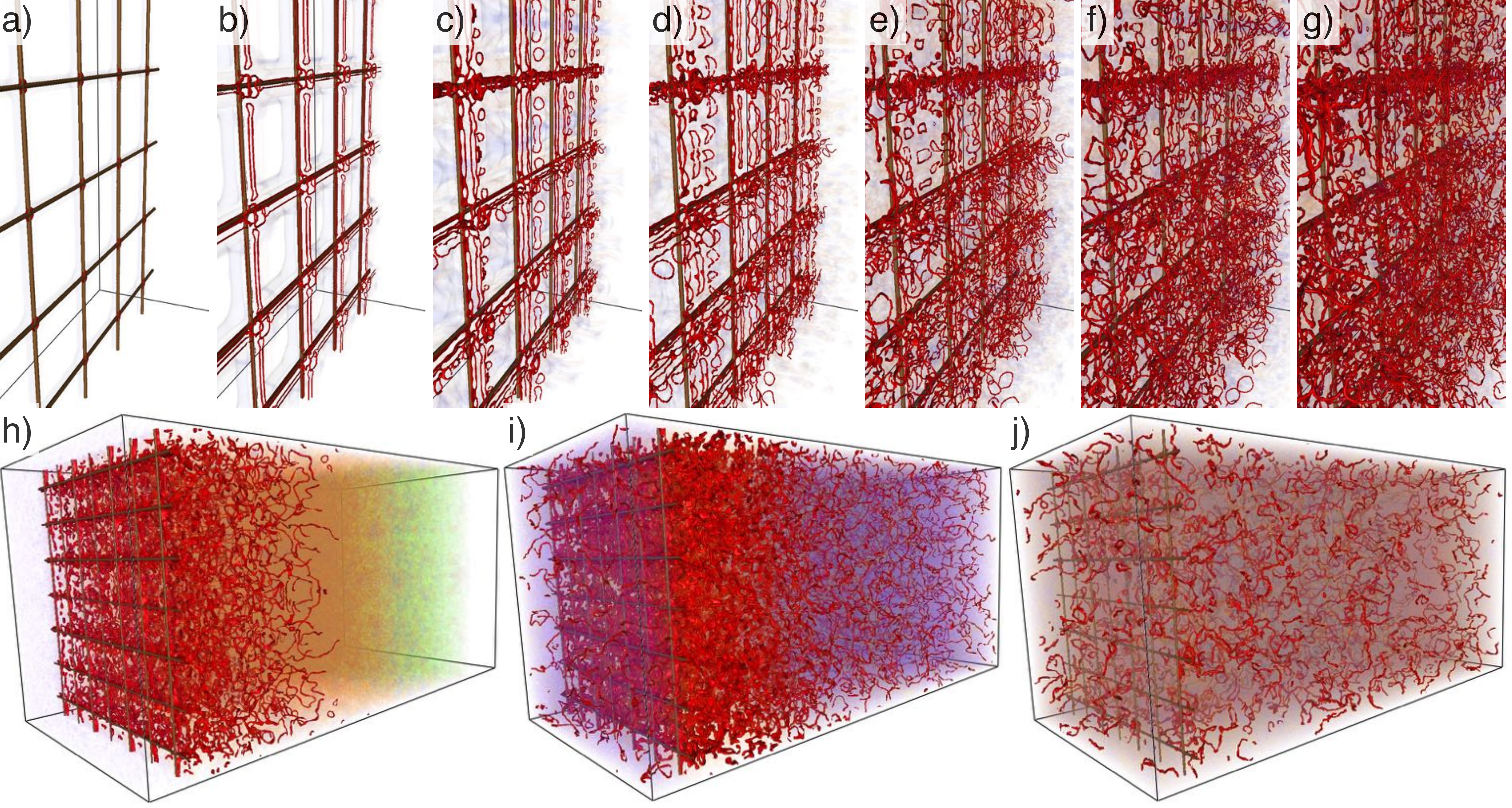}
\caption{(Color online)  3D visualizations of the density field (rendered with the software VAPOR). Red isosurfaces are at low density values corresponding to vortices. Blue and green density clouds correspond to sound waves (density fluctuations around $\rho=1$). a-g) Zoom close to the grid at early times $t=0.1,\,0.6,\,1.1,\,1.6,\,2.4,\,3.1$ and $6.4$ (from left to right) for run c1. h-j) Run b3 at $t=21,70,126$.}
\label{Fig:3DSnapshotsGrid}
\end{figure*}

{We first focus on simulations done using the grid.} The grid generates a turbulent wake that is displayed in Figs.\ref{Fig:3DSnapshotsGrid}.a-d by the isosurface of the density.
At early times vortices are nucleated close to the grid (Figs.\ref{Fig:3DSnapshotsGrid}.a-c, run c1), leading later to a turbulent wake (Figs.\ref{Fig:3DSnapshotsGrid}.h-j, run b3). It is well known in the {framework} of $2D$ GP that vortex dipoles are nucleated {behind a cylindrical obstacle} for Mach numbers above a {critical} threshold $M_{\rm crit}\approx 0.4$ \cite{huepe2000nucleation}. The equivalent in $3D$ are vortex rings that rapidly reconnect and create the complex tangle observed in Fig.\ref{Fig:3DSnapshotsGrid}. The process is identical to the one described for $^3$He-B experiments {using a grid reported } in \cite{Bradley2005}. 

\begin{figure}[h!]
\includegraphics[width=0.485\textwidth]{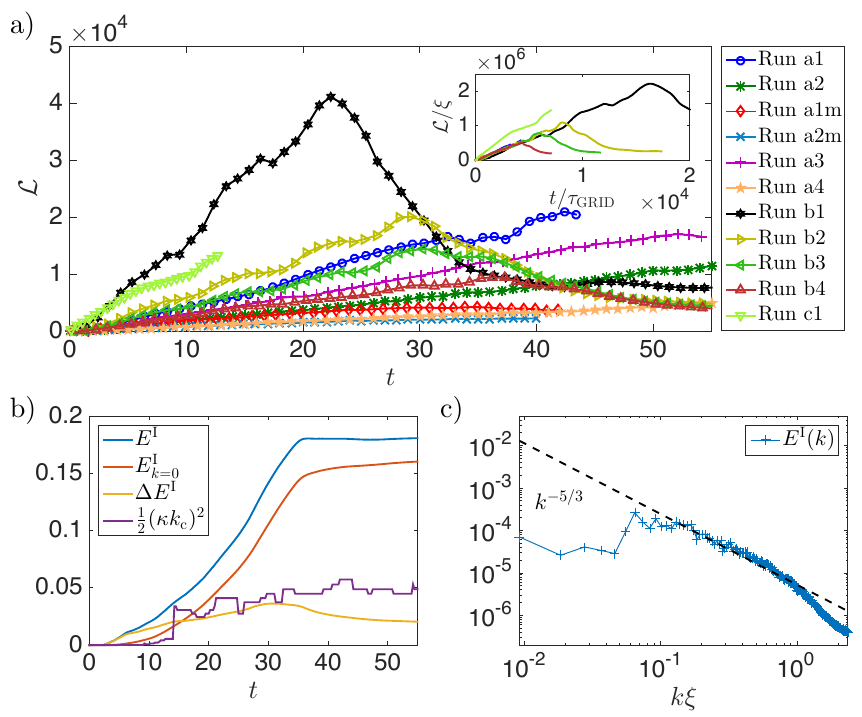}
\caption{(Color online) a) Total vortex line for different grid runs. The inset shows the collapse of $\mathcal{L}/\xi$ vs $t/\tau_{\rm Grid}$, with $\tau_{\rm Grid}=T_{\rm nucl}D\xi /L_\perp^2$ (see text).  b) Temporal evolution of $E^{\rm I}$, the kinetic energy $E^{\rm I}_{k=0}$ of the mean incompressible velocity field, the turbulent fluctuations $\Delta E^{\rm I}=E^{\rm I}-E^{\rm I}_{k=0}$ and the kinetic energy of the condensate (all energies averaged over $z$). Run b1. c) Averaged energy spectrum for run c1 at $z/\xi\in(355,365)$ and $t=5.5$.}
\label{Fig:GridGlobal}
\end{figure}

Two stages are observed during the development of turbulence in the wake of the grid. During the first stage, incompressible kinetic energy is injected by nucleation of rings. To account for this, the total vortex length is measured as $\mathcal{L}(t)=\int \theta(0.2- \rho({\bf x},t))d^3\,\x/\int \theta(0.2- \rho^{\rm 2D}({\bf x}))d^2\x$, where $\theta(\,)$ is the Heaviside function and $\rho^{\rm 2D}({\bf x})$ is the profile of a two-dimensional vortex given by the Pad\'e approximation \cite{pismen1999vortices}. Note that $\mathcal{L}(t)$ is only a rough estimate as small amplitude Kelvin waves and density oscillations along filaments modify this volume integral. The temporal evolution of $\mathcal{L}(t)$ is displayed in Fig.\ref{Fig:GridGlobal}.a for all runs. 
The increase of the vortex length depends on the geometry of the grid and on the Mach number. From Fig.\ref{Fig:3DSnapshotsGrid}.a-g we observe two kind of structures: large elongated rings of length $\sim 2D$ and smaller ones close to the corners. When $\xi\ll D$ we expect that the main contribution to $\mathcal{L}$ comes from elongated rings. The vortex length can be thus estimated as $\mathcal{L}(t)\sim D (L_\perp/D)^2\,t/T_{\rm nucl}$, where $T_{\rm nucl}$ is the typical time between two nucleations ($(L_\perp/D)^2$ is proportional to the number of such rings). The characteristic time-scale of vortex injection for the grid is thus defined in terms of $T_{\rm nucl}$ as $\tau_{\rm Grid}=T_{\rm nucl}D\xi /L_\perp^2$ (so that $\mathcal{L}/\xi\sim t/\tau_{\rm Grid}$). Vortex nucleation in a superflow around a disc has been extensively studied in the last 15 years \cite{frisch1992transition,huepe2000nucleation,SasakiVonKarmanGP2d}. It was found that stable and unstable (nucleation) branches are connected through a primary saddle-node and a secondary pitchfork bifurcation. The critical Mach number $M_{\rm crit}$ was found to depend on $a/\xi$ and close to the critical point $T_{\rm nucl}$ to scale as $T_{\rm nucl}\sim (M/M_{\rm crit}-1)^{-1/2}$, that corresponds to a dissipative saddle-node bifurcation. For the grid, geometry is more complex and  $M\gg M_{\rm crit}$, so the previous scaling is not expected to be valid. A precise determination of $T_{\rm nucl}$ is out of the scope of the present work. However it can be empirically observed that $T_{\rm nucl}\sim (\xi/c)(D/\xi)^{1/2} M^{-4}$ is compatible with data presented in this {work}. This is manifested by the relatively good collapse of $\mathcal{L}/\xi$ for the different runs displayed in the inset of Fig.\ref{Fig:GridGlobal}.a. A precise study of the nucleation will be performed in a future work.

Figure \ref{Fig:GridGlobal}.b displays the temporal evolution of the incompressible kinetic energy of run b1. The saturation of the energy is related to the growth of the mean flow, as shown by the temporal evolution of the $k=0$ mode $E^{\rm I}_{k=0}$ of the $3D$ energy spectrum (including the average over $z$). Note that the energy fluctuations $\Delta E^{\rm I}=E^{\rm I}-E^{\rm I}_{k=0}$ reaches a maximum and then decreases, consistently with the decay of $\mathcal{L}(t)$ in Fig.\ref{Fig:GridGlobal}.a. The time when the vortex length and energy fluctuations are the largest, corresponds to the time when the bulk of vortices reaches the opposite side of the box (respect to the grid). This fact has been checked with runs using the same parameters but with larger $L_\parallel$ (data not shown). By this time, the condensate (initially at the wavenumber ${\bf k_{\rm c}=0}$) has ``jumped'' to higher wavenumbers. This can be interpreted as the full system being entrained by the imposed flow.
Indeed, a boost of velocity ${\v_{\rm G}}$ corresponds for GP to a multiplication of $\psi$ by $e^{i\frac{m}{\hbar}{\v_{\rm G}}\cdot \k}$ \cite{Krstulovic2011b}. The mean (incompressible) kinetic energy of the condensate is thus given by $\frac{1}{2} (\hbar/m)^2|{\bf k_{\rm c}}|^2$, where ${\bf k_{\rm c}}$ is determined by the wavenumber with the largest number of particles. The temporal evolution of this energy is also shown in Fig.\ref{Fig:GridGlobal}.b. The full system moving at velocity $k_{\rm c} \hbar/m$ has an effective Mach number $(v_0-k_{\rm c} \hbar/m)/c$  (e.g. $\sim 4.6$ for run b1). At this Mach number ring nucleation stops, leading to the later decay of $\mathcal{L}(t)$ (runs b1-b4). For all other runs, the integration is not long enough to observe the decay. The same phenomenon is observed in equivalent $2D$ simulations (as the one in \cite{Reeves2013}) if the integration is performed for longer times (data not shown). 

Before the decay starts, a turbulent state is observed. The energy spectrum computed at a distance $z\approx360\xi$ from the grid is displayed in Fig.\ref{Fig:GridGlobal}.c. 
A Kolmogorov scaling is expected to be observed for  $k\ll k_{\rm IV}=2\pi/\ell_{\rm IV}$, where $\ell_{\rm IV}$ is the inter-vortex distance, usually estimated as $\ell_{\rm IV}=1/\sqrt{\mathcal{L}/V}$ . For the grid, turbulence is not homogeneous, but an effective volume can be obtained through a spatial average weighted by $\Delta E^{\rm I}(z)=\sum_{k>0}E^{\rm I}_k(z)$.
A plot of $\Delta  E^{\rm I}(z)$ and details on this average are included in the {Appendix \ref{Appendix:Intervortex}}. For the corresponding run and time of Fig.\ref{Fig:GridGlobal}.c we obtain $V_{\rm eff}=L_\perp^2\times 169\xi$ that yields $\ell_{\rm IV}=10.6\xi$. This corresponds to $ k_{\rm IV}\xi\approx.6$, which is in good agreement with the end of the $k^{-5/3}$ scaling observed in Fig.\ref{Fig:GridGlobal}.c. 
Finally, in the second stage, rings shrink due to mutual friction effects \cite{Krstulovic2011b}. The estimations of $\mathcal{L}$ by a volume integral does not allow us to verify the Vinen's decay prediction \cite{Vinen493}.

\begin{figure}[h!]
\includegraphics[width=0.45\textwidth]{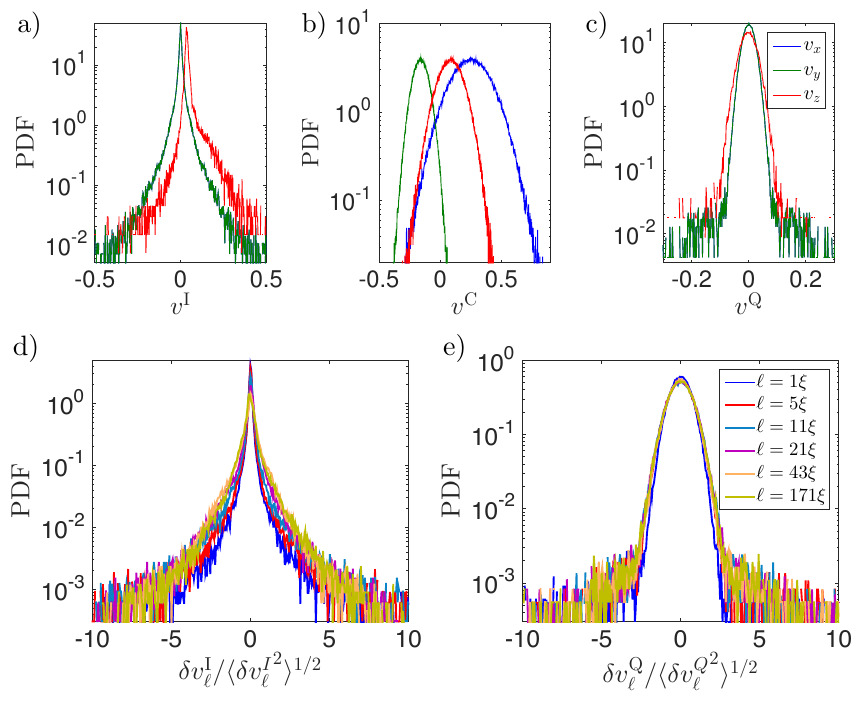}
\caption{(Color online) a-c) One-point velocity PDF of the three components of ${\bf v}^{\rm I}$, ${\bf v}^{\rm C}$ and ${\bf v}^{\rm Q}$ (same color code for a-c). d-e) PDFs of the velocity increments of $\delta v^I/\delta v^I_{\rm vrms}$ and $\delta v^Q/\delta v^Q_{\rm vrms}$ for different increments $\ell$. Same scales as in e).
All the data from run c1 taken at $t=5.5$ and $z=360\xi$. }
\label{Fig:GridStat}
\end{figure}

One the most remarkable differences between classical and superfluid turbulence is the one-point velocity statistics \cite{Baggaley2011a,LaMantia2014,White2010}. The probability distribution function (PDF) of $\v=\nabla\phi$ presents power law tails $\sim v^{-3}$ unlike the Gaussian PDFs observed in the classical case. As in GP, the velocity field $\v$ is ill-defined at the vortex core because the phase is not defined, we instead look at statistics of the regularized fields \eqref{Eq:Velos}. Non-Gaussian PDF have already been observed for ${\bf v}^{\rm I}$ in $2D$ GP simulations \cite{Shukla2013}. The PDFs of the three components of the velocity are displayed in Fig.\ref{Fig:GridStat}a-c for ${\bf v}^{\rm I}$, ${\bf v}^{\rm C}$ and ${\bf v}^{\rm Q}$ at a fixed time and distance {from} the grid for run c1.
The PDFs of ${\bf v}^{\rm I}$ present as in reference \cite{Shukla2013} non-Gaussian tails scaling as $ {v^{\rm I}}^{-b}$ with $b\in (2,3)$. On the contrary ${\bf v}^{\rm C}$ and ${\bf v}^{\rm Q}$ exhibit almost Gaussian PDFs. These can be explained because sound waves (related to ${\bf v}^{\rm C}$ and ${\bf v}^{\rm Q}$) are indeed expected to thermalize at small scales and thus to develop Gaussian statistics \cite{Krstulovic2011a}. 

\begin{figure}
\includegraphics[width=0.45\textwidth]{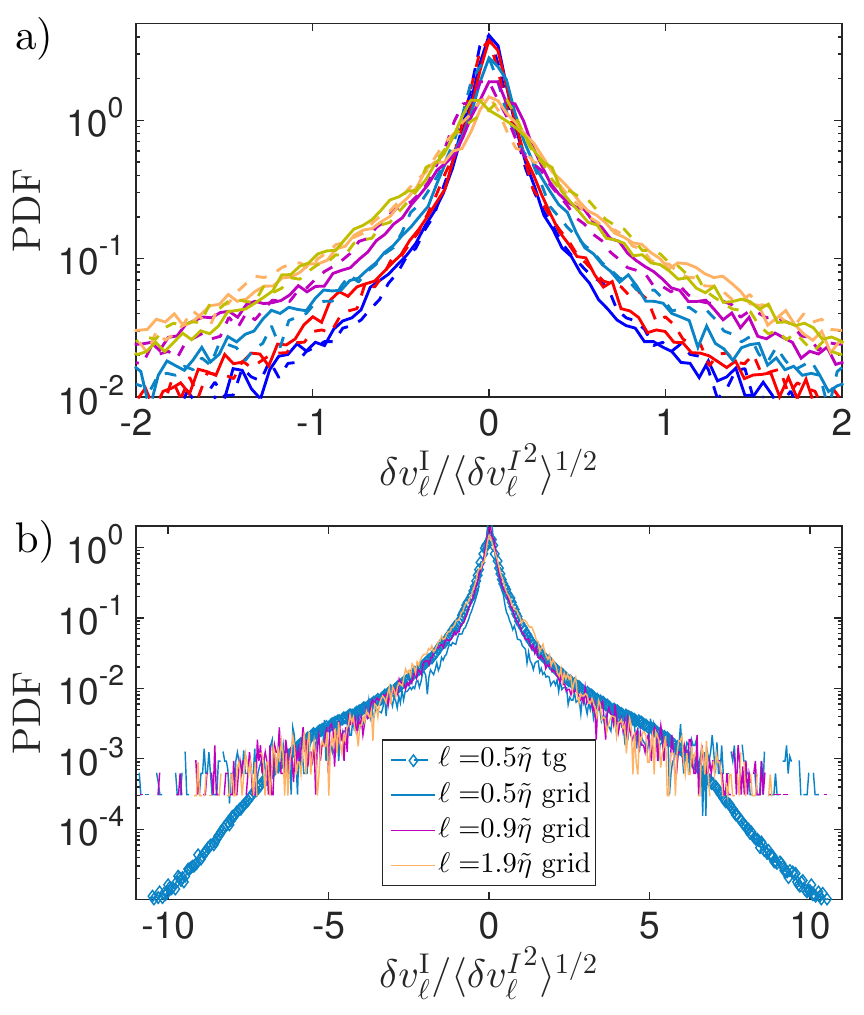}
\caption{(Color online) a) Zoom of Fig\ref{Fig:GridStat}.a together with the corresponding PDFs of $-\delta v_\ell  ^{\rm I}/\langle {\delta v_\ell  ^{\rm I}}^2\rangle^{1/2}$ in dashed lines. b) Comparison between grid (run c1) and Taylor-Green turbulence.}
\label{Fig:GridStatANDCompare}
\end{figure}

Motivated by classical turbulence we define the longitudinal velocity increments as
\begin{equation}
\delta v_\ell  ^{\rm \Lambda}=({\bf v}^{\rm \Lambda}({\bf x}+\ell \hat{{\bf r}})-{\bf v}^{\Lambda}({\bf x}))\cdot \hat{{\bf r}}
\end{equation} 
with $\hat{r}$ a unit vector. For the grid $\hat{r}$ is taken perpendicular to the mean flow. The velocity increments {PDFs} are displayed in Fig.\ref{Fig:GridStat}d-e for I and Q. As in classical turbulence $\delta v_\ell  ^{\rm I}/\langle {\delta v_\ell  ^{\rm I}}^2\rangle^{1/2}$ presents strongly non-Gaussian statistics that depend on the scale $\ell$, manifesting the non self-similar behavior of turbulence. 
Figure \ref{Fig:GridStatANDCompare}.a shows a zoom of Fig.\ref{Fig:GridStat}d together with the PDF of $-\delta v_\ell  ^{\rm I}/\langle {\delta v_\ell  ^{\rm I}}^2\rangle^{1/2}$ in dashed lines. A negative skewness is apparent {there}.
This asymmetry of the PDFs is an important property of Kolmogorov turbulence related to the energy cascade and the $4/5$-law of turbulence \cite{Frisch1995}. The bulk of the PDF of $\delta v_\ell  ^{\rm Q}$ is Gaussian whereas the tails depend on the scale and separate from Gaussian statistics. No skewness is observed. The increment of $\delta v_\ell  ^{\rm C}$ are totally Gaussian and scale independent once normalized by their rms value (not shown).

It is well known that strong velocity fluctuations in classical turbulence lead to the breakdown of the totally self-similar Kolmogorov phenomenology (K41). Intermittency is responsible {for} this breakdown and it is quantified by looking at the scaling of the velocity increments moments
\begin{equation}
S^{\rm I}_p(\ell)=\langle ||({\bf v}^{\rm I}({\bf x}+\ell \hat{{\bf r}})-{\bf v}^{\rm I}({\bf x}))||^p\rangle,\label{Eq:Sp}
\end{equation}
known as structure functions (average is over all directions of $\hat{{\bf r}}$). In the inertial range, i.e. at scales smaller than the integral scale $L_{\rm int}$ and larger than the dissipative scale $\eta$, it is expected that $S^{\rm I}_p(\ell)\sim\ell^{\zeta_p^{\rm I}}$. K41 predicts $\zeta_p=p/3$, whereas numerical and experimental results evidence a non-linear function \cite{Frisch1995}. The deviation from $\zeta_p=p/3$ are known as intermittency corrections and $\zeta_p$ as anomalous exponents. Note the (analytical) $4/5$-law fixes $\zeta_3=1$. We now address this issue within the framework of GP. The statistics presented in Fig.\ref{Fig:GridStat} do not allow to obtain a clear scaling. In order to obtain a larger inertial range {and better statistics}, we make use of the Taylor-Green flow. This flow is known to develop a vortex tangle with a $k^{-5/3}$ energy spectrum \cite{Nore1997a}. A visualization of the {Taylor-Green flow is presented in Fig.\ref{Fig:3DSnapshotsTG} at different times}.
\begin{figure*}
\includegraphics[width=0.9\textwidth]{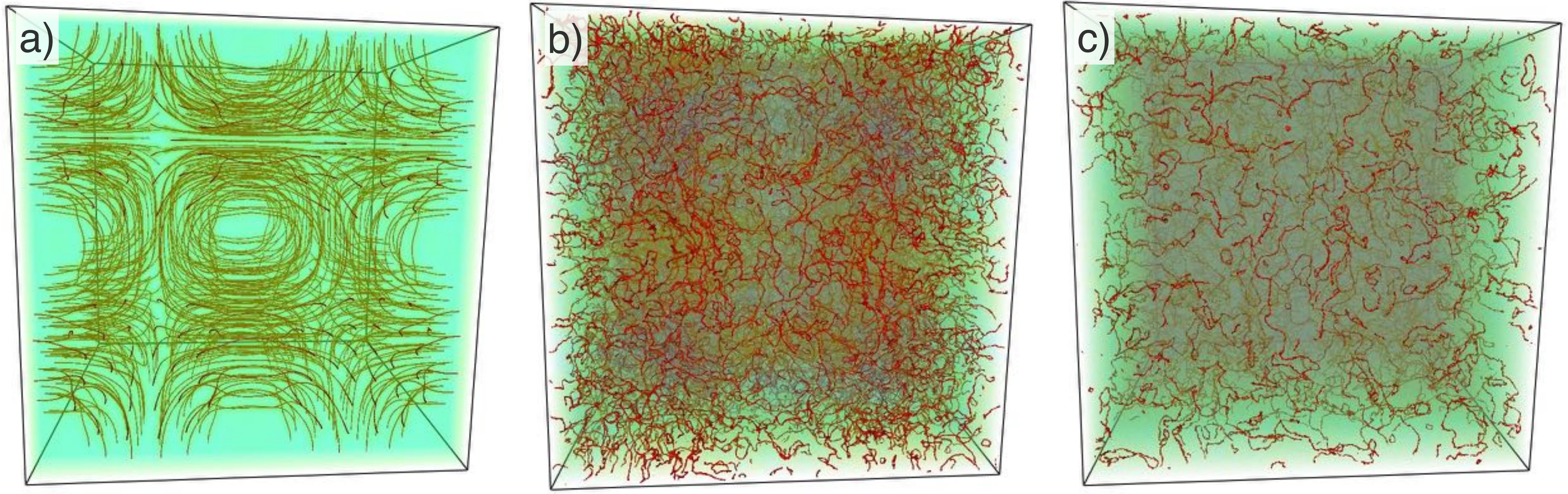}
\caption{(Color online) (Color online)  3D visualizations of the density field (rendered with the software VAPOR). Red isosurfaces are at low density values corresponding to vortices. Green density clouds correspond to sound waves (density fluctuations around $\rho=1$). Temporal evolution (left to right) of Taylor-Green vortex. $t=0,12.5,33.5$.}
\label{Fig:3DSnapshotsTG}
\end{figure*}
We first compare the statistics of the TG velocity increments with those of the grid. 
We define a Kolmogorov (like) dissipative scale $\tilde{\eta}$ using the quantum of circulation as $\tilde{\eta}=L_{\rm int} (v_{\rm rms}^{\rm I}L_{\rm int} /\kappa)^{-3/4}$, where the integral scale $L_{\rm int}$ is estimated as $D$ and $L/2$ for the grid and the Taylor-Green flow respectively. The corresponding values are $\tilde{\eta}_{\rm Grid}=23\xi$ for run c1 and $\tilde{\eta}_{\rm TG}=8.5\xi$ for Taylor-Green run. Note in Fig.\ref{Fig:GridStatANDCompare}.b that the statistics of both flows coincide, if velocity increments are compared at similar scales (in units of $\tilde{\eta}$). This is a manifestation universality in quantum turbulence like the one observed in classical turbulence. Slight discrepancies between the two configurations and small values of $\ell/\tilde{\eta}$ are due to the non unique way of defining a Kolmogorov length in quantum turbulence. 
 
\begin{figure}[h!]
\includegraphics[width=0.47\textwidth]{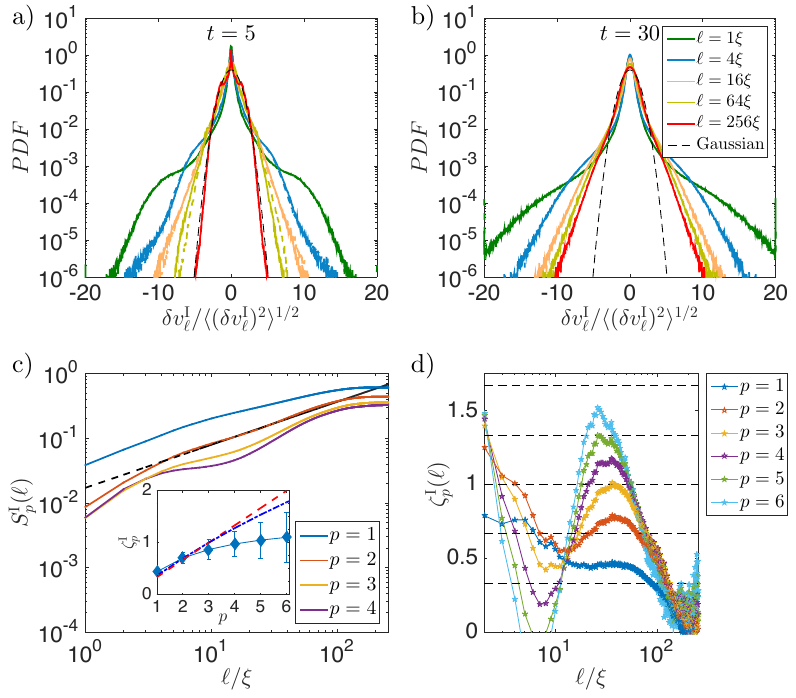}
\caption{(Color online) Taylor-Green run. a) PDFs of the velocity increments at $t=5$ for different scales (colors as in (b)). Dashed lines correspond to PDFs of $-\delta v_\ell^I /\langle (\delta v_\ell^I)^2\rangle^{1/2}$. b) Idem as a) but for $t=30$. c) Structure functions \eqref{Eq:Sp} at $t=5$. The black dashed line represent the K41 prediction $S_2^{\rm I}(\ell)\approx (11/3)C_2 (\epsilon \ell) ^{2/3}$ with $C_2=2.6625$ and $\epsilon=-\frac{dE^{\rm I}}{dt}$. The inset displays the anomalous exponent $\zeta_p^{\rm I}$, the red dashed line the Kolmogorov scaling $\zeta_p=p/3$ and the point-dashed blue line the She-L\'ev\^eque model. d) Local slope of the structure functions. 
\label{Fig:TG}}
\end{figure}

The PDFs of Taylor-Green flow velocity increments at different scales are displayed in Fig.\ref{Fig:TG}.a-b for two different times.
At $t=5$ a clear scale dependence and skewness are observed. {For instance}, the skewness is equal to $-0.13$ for $\ell=16\xi$ at $t=5$. At later times ($t=30$), the increments tend towards non-skewed PDFs, thought not Gaussian. The structure functions are presented in Fig.\ref{Fig:TG}.c. K41 predicts $S^{\rm I}_2(\ell)\approx (11/3)C_2 (\epsilon \ell)^{2/3}$, with $\epsilon$ the energy dissipation rate and $C_2$ the Kolmogorov constant \cite{pope2000turbulent}. In GP, $\epsilon$ can be estimated as $\epsilon=-\tfrac{dE^{\rm I}}{dt}$. By using this estimation we obtain {(by fitting)} $C_2=2.6625$ (see dashed line in Fig.\ref{Fig:TG}.c). Note that $C_2=2.6625$, is very close to the value $2.0\pm0.4$ reported in classical turbulence.

To look at the intermittency, the local slopes $\zeta_p(\ell)=\tfrac{d\log{S^{\rm I}_p(\ell)}}{d\log{\ell}}$ are presented in Fig.\ref{Fig:TG}.d. {A power-law scaling of $S^{\rm I}_p(\ell)$ corresponds to the plateau in $\zeta_p(\ell)$.} The $\zeta_p^{\rm I}$ are measured averaging the local slope for $\ell/\xi\in(15,75)$. The anomalous exponents are displayed in the inset of Fig.\ref{Fig:TG}.c. The red dashed and blue dot-dashed lines represents K41 and She-L\'ev\^eque model respectively \cite{she1994universal}. GP intermittency is found to be stronger than the classical one (that is in general well represented by She-L\'ev\^eque model). Intermittency was already measured in $^4$He by early experiments performed by Maurer et al.\cite{maurer1998local} and no difference with classical experiment was found. However, it has been observed that intermittency of the von K\'arm\'an flow (in classical fluids) is slightly stronger than other turbulent flows \cite{salort2011investigation}. As the Taylor-Green flow mimics the von K\'arm\'an flow, an enhancement of intermittency could also be expected. 
In addition, using HVBK-based shell models \cite{boue2013enhancement,shukla2015multiscaling} a clear temperature dependence has been observed for the $\zeta_p$, presenting a maximum of intermittency around $0.6T_\lambda$ (with $T_\lambda$ the temperature of the $\lambda$-point).  These HVBK results do not directly apply to GP turbulence, that formally describes the low temperature limit of BECs. Indeed, in this limit dissipation need to be added by some ad-hoc mechanism to the HVBK model, unlike GP, where energy of vortices is naturally dissipated by phonon radiation. Furthermore, in the HVBK there is no notion of quantized vortices, as only a large-scale description is given. The results presented in this {work} directly apply to BECs at low temperature but are also expected to be relevant for superfluid Helium. Although today {it} is possible to create and track several vortex lines in BECs \cite{Serafini2015_VortexBEC,Lamporesi2013VortexCreation}, a controlled experiment with such a dense turbulent vortex tangle is not still realizable. However, the large fluctuations of velocity fields reported in this {work} are expected to be an inherent property of turbulent BECs that could be observed in the future. 

Understanding of intermittency in classical flows remains an open problem, in superfluids not enough information is available. The simulations presented here are not in a statistically steady-state which is the most suitable configuration for such a study. However, it has been shown that velocity statistics of grid turbulence are similar to these of Taylor-Green. Grid simulations could be thus used to investigate intermittency if the injection/dissipation is modified to obtain a stationary regime. Much longer simulations at higher resolutions are needed.

\begin{acknowledgments}  
The author acknowledges useful scientific discussions with J. Bec, M.E. Brachet, V. Shukla.
Computations were carried out at M\'esocentre SIGAMM hosted at the Observatoire
de la C\^ote d'Azur.
\end{acknowledgments}

\appendix
\section{Model and procedure\label{Appendix:Model}}

\begin{figure*}
\includegraphics[width=0.9\textwidth]{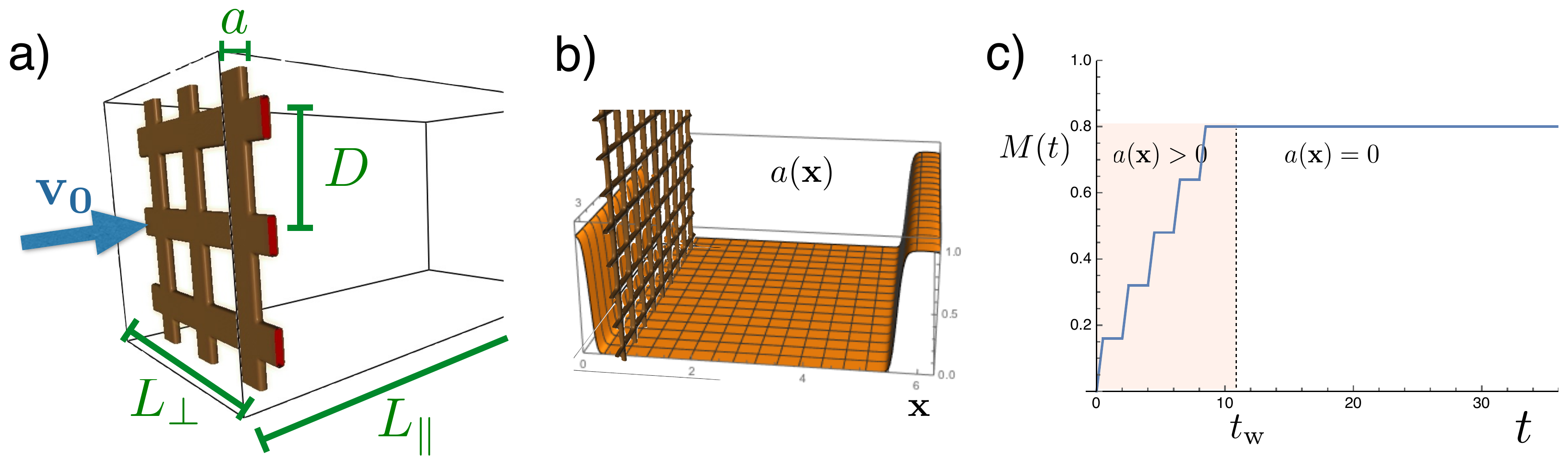}
\caption{(Color online) a) Scheme of the domain. b) Scheme of local dissipation. c) Protocol followed to increase the Mach number (arbitrary units).}
\label{Fig:Grid}
\end{figure*}

We consider the Gross-Pitaevskii equation. The grid is modelled by a strong repulsive potential $V_{\rm grid}({\bf x})$ and an advection term is added in the left hand side to impose the mean flow. The final equation reads:
\begin{equation}
i\hbar\left(\frac{\partial \psi}{\partial t}+\vec{v_0}\cdot\nabla\psi\right)=- \frac{\hbar^2}{2m}\nabla^2 \psi + g\,|\psi|^2\psi -\tilde{\mu}\, \psi+V_{\rm grid}({\bf x})\,\psi.\label{Eq:GPadvGrid}
\end{equation}
When $\vec{v_0}=0$ and $V_{\rm grid}({\bf x})=0$, Eq.\eqref{Eq:GPadvGrid} conserves the total energy $H=\int (\alps|\nabla \psi|^2 +\frac{g}{2}|\psi|^4)d\x$ and the total number of particles $\int |\psi|^2d\x$.
The grid potential $V_{\rm grid}({\bf x})$ is defined as follow:
\begin{eqnarray}
&&V_{\rm grid  }(x,y,z)=V_0\, V_{\rm rod 1D}(z,z_0,a_\parallel,L_\parallel)V_{\rm grid 2D}(x,y)\\
\nonumber && V_{\rm grid 2D}(x,y)= \min{[V_{\rm grid 1D}(x)+V_{\rm grid 1D}(y),1]}\\
\nonumber && V_{\rm grid 1D}(x)=\sum_{i=1}^{L_\perp/D} V_{\rm rod 1D}(x,(i-\tfrac{1}{2})D,a_\perp,L_\perp)\\
\nonumber && V_{\rm rod 1D}(x,x_0,a,L) =\frac{1}{2}+\frac{1}{2}\tanh{\left[\frac{\cos^2 {\tfrac{\pi (x-x_0)}{L}}-\cos^2{(\tfrac{\pi a}{2L})}}{\epsilon^2/2^2}\right]}.\label{Eq:gridPot}
\end{eqnarray}
The distance between the rods of the grid is given by the mesh size $D$, the diameter of each rod of the grid is given by $a_\parallel$ and $a_\perp$ in the parallel and perpendicular direction respectively (respect to the flow). The dimensions of the box are $L_\perp\times L_\perp\times L_\parallel$. $z_0$ is the position of the grid. See Fig.\ref{Fig:Grid}.a for an illustration of the grid and the box. For all simulations $a_\parallel=a_\perp=2\xi$, $\epsilon=.5\xi$, $z_0=L_\perp/8$ and $V_0=20c^2$. See Table \ref{Table:RUNS} for the values of the other parameters and the list of the different runs. 

The initial condition containing the grid and the fluid at rest is obtained by using a Newton-Raphson method that ensures a perfect and clean initial condition \cite{huepe2000nucleation}. However, if Eq.\ref{Eq:GPadvGrid} is abruptelly integrated with $v_0\neq0$ a lot of sound is emitted. In order to minimize the initial emission of sound a local dissipative version of Eq.\ref{Eq:GPadvGrid} is used for the short times, typically up to $t_{\rm W}\sim L_\parallel/c$ that corresponds to the time of sound waves take to travers the box. The dissipative GP is a mix of the Real Ginzburg-Landau and GPE, namely
\begin{widetext}
\begin{equation}
i\hbar\left(\frac{\partial \psi}{\partial t}+\vec{v_0}\cdot\nabla\psi\right)=(1-i a({\bf x}))\left(- \frac{\hbar^2}{2m}\nabla^2 \psi + g\,|\psi|^2\psi -\tilde{\mu}\, \psi+V_{\rm grid}({\bf x})\,\psi\right),\label{Eq:GPadvGridDisp}
\end{equation}
\end{widetext}
where $a({\bf x})$ is zero almost everywhere but close to the faces opposite to the grid, as displayed in Fig.\ref{Fig:Grid}.b. Furthermore, the Mach number $M=v_0/c$ is increased by steps as schematized in Fig.\ref{Fig:Grid}.c. After $t_{\rm W}$ only the advective GP is used.

\section{Numerical integration\label{Appendix:Numerics}}

We use a standard pseudo-spectral code with Runge-Kutta of order 2 for time stepping. Note that the potential in Eq.\eqref{Eq:gridPot} has been carefully chosen periodic. To have a fully de-aliased code, the scheme proposed in reference \cite{Krstulovic2011b} is used (see Appendix of that reference). It consists on applying the Galerkin projector $\mathcal{P}_{\rm G}$ as follows
\begin{eqnarray}
\nonumber i\hbar\left(\frac{\partial \psi}{\partial t}+\vec{v_0}\cdot\nabla\psi\right)&=&\mathcal{P}_{\rm G}\bigg[- \frac{\hbar^2}{2m}\nabla^2 \psi + g\,\mathcal{P}_{\rm G}\left[|\psi|^2\right]\psi  \\
\nonumber&&\hspace{2.25cm}-\tilde{\mu}\, \psi+V_{\rm grid}({\bf x})\,\psi\bigg].\\
\label{Eq:GPadvGridProj}
\end{eqnarray}
%\end{widetext}
The Galerkin projector $\mathcal{P}_{\rm G}$ takes a simple form in Fourier space: $\mathcal{P}_{\rm G}[\hat{\psi({\bf k})}]=\theta(\kmax-|k|)\hat{\psi({\bf k})}$ where $\theta$ is the Heaviside function and $\kmax=N/3$ is chosen following the standard $2/3$ rule for a quadratic non-linearity \cite{gottlieb1977numerical}. When de-aliasing is not performed as in \eqref{Eq:GPadvGridProj}, conservation of momentum is not preserved by the discrete system. For the grid simulation, the system is not isotropic and it has finite momentum in one direction. The lack of momentum conservation typically leads to spurious and non controlled effects. The additional projector applied in Eq.\eqref{Eq:GPadvGridProj} costs one extra back and forth FFT per time step. An alternative to such technique is to use the standard de-aliasing but with  $\kmax=N/4$ (corresponding to a cubic non-linearity) at the price of wasting half of the resolution.

The Taylor-Green flow is prepared as in reference \cite{Nore1997} (with a different choice of parameters) but no symmetries are imposed during the temporal evolution. Symmetries are thus not preserved for all times.

\section{Inter-vortex distance\label{Appendix:Intervortex}}

\begin{figure}
\includegraphics[width=0.45\textwidth]{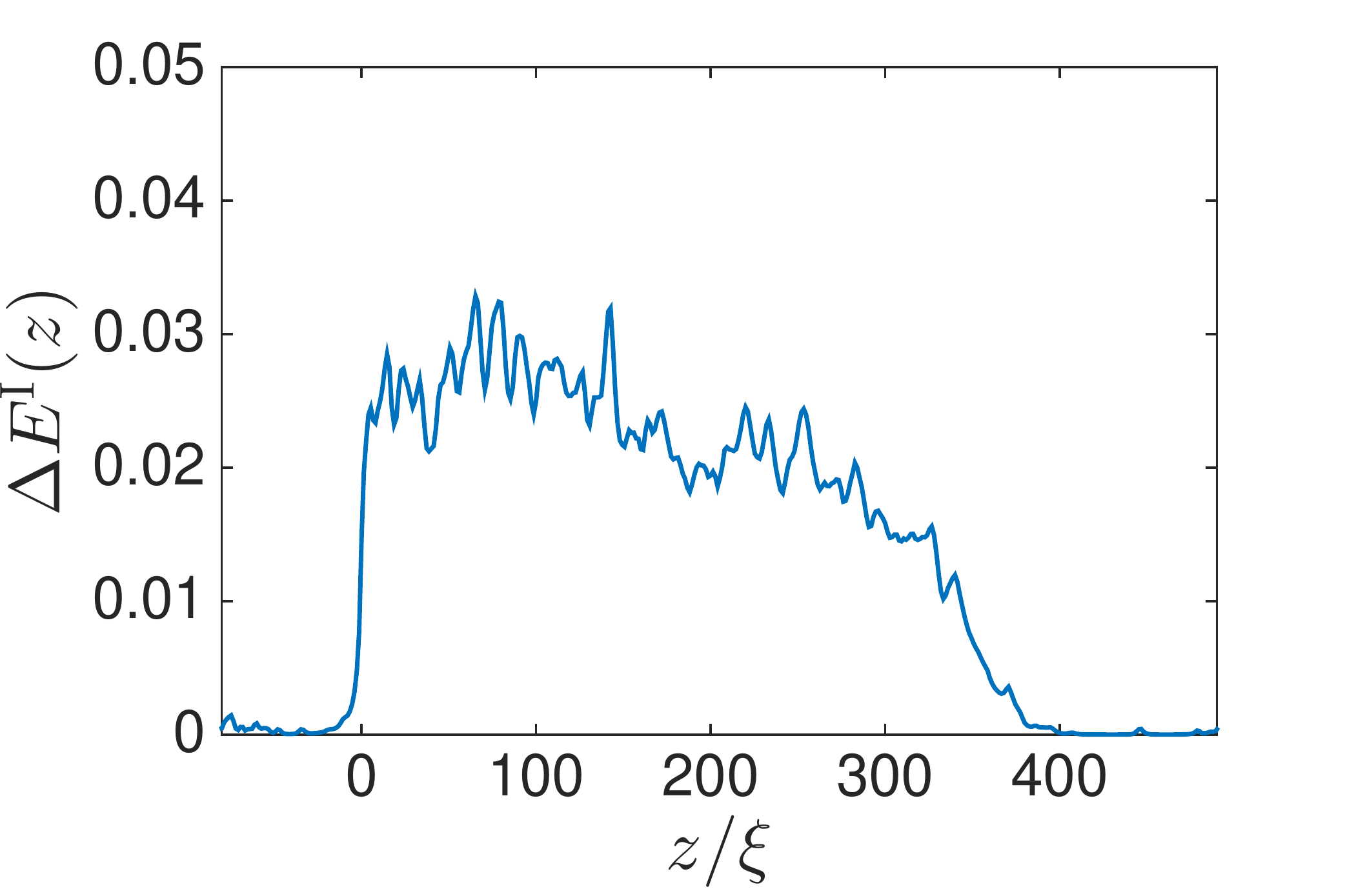}
\caption{(Color online) Spatial dependence of the incompressible kinetic energy $\Delta E^{\rm I}(z)$. Run c1 at t=5.5}
\label{Fig:DeltaEz}
\end{figure}

The inter-vortex distance is usually estimated as $\ell_{\rm IV}=1/\sqrt{\mathcal{L}/V}$, where $\mathcal{L}$ is the total vortex length and $V$ the volume of the system. For the grid, turbulence is not homogenous therefore an estimation of the volume is needed. Vortices manly contribute to the incompressible kinetic energy. This quantity can be thus used to estimate the size of the turbulent bulk. We define energy fluctuations due to turbulence as a function of the distance to the grid $z$ by
\begin{equation}
\Delta E^{\rm I}(z)=\sum_{k>0} E^{\rm I}_k( z),
\end{equation}
where $E^{\rm I}_k( z)$ is the energy spectrum computed with the wavevectors perpendicular to the mean flow for a fixed value $z$. This quantity is displayed in Fig.\ref{Fig:DeltaEz} for run c1 at $t=5.5$. The bulk is clearly visible.
The effective volume containing the bulk is defined as $V_{\rm eff}=L_\perp^2\times L_{\rm eff}$ with $L_{\rm eff}=\left(\int_0^{L_\parallel} z \Delta E^{\rm I}(z) \mathrm{d}z\right)/\left(\int_0^{L_\parallel} \Delta E^{\rm I}(z) \mathrm{d}z\right)$.
\begin{equation}
L_{\rm eff}=\frac{\int_0^{L_\parallel} z \Delta E^{\rm I}(z) \mathrm{d}z}{\int_0^{L_\parallel} \Delta E^{\rm I}(z) \mathrm{d}z}.
\end{equation}
Finally, the inter-vortex distance is estimated as $\ell_{\rm IV}=\sqrt{V_{\rm eff}/\mathcal{L}}$.%$\ell_{\rm IV}=1/\sqrt{\mathcal{L}/V_{\rm eff}}$.

\bibliographystyle{apsrev4-1}
%\bibliography{Grid}

\begin{thebibliography}{33}%
\makeatletter
\providecommand \@ifxundefined [1]{%
 \@ifx{#1\undefined}
}%
\providecommand \@ifnum [1]{%
 \ifnum #1\expandafter \@firstoftwo
 \else \expandafter \@secondoftwo
 \fi
}%
\providecommand \@ifx [1]{%
 \ifx #1\expandafter \@firstoftwo
 \else \expandafter \@secondoftwo
 \fi
}%
\providecommand \natexlab [1]{#1}%
\providecommand \enquote  [1]{``#1''}%
\providecommand \bibnamefont  [1]{#1}%
\providecommand \bibfnamefont [1]{#1}%
\providecommand \citenamefont [1]{#1}%
\providecommand \href@noop [0]{\@secondoftwo}%
\providecommand \href [0]{\begingroup \@sanitize@url \@href}%
\providecommand \@href[1]{\@@startlink{#1}\@@href}%
\providecommand \@@href[1]{\endgroup#1\@@endlink}%
\providecommand \@sanitize@url [0]{\catcode `\\12\catcode `\$12\catcode
  `\&12\catcode `\#12\catcode `\^12\catcode `\_12\catcode `\%12\relax}%
\providecommand \@@startlink[1]{}%
\providecommand \@@endlink[0]{}%
\providecommand \url  [0]{\begingroup\@sanitize@url \@url }%
\providecommand \@url [1]{\endgroup\@href {#1}{\urlprefix }}%
\providecommand \urlprefix  [0]{URL }%
\providecommand \Eprint [0]{\href }%
\providecommand \doibase [0]{http://dx.doi.org/}%
\providecommand \selectlanguage [0]{\@gobble}%
\providecommand \bibinfo  [0]{\@secondoftwo}%
\providecommand \bibfield  [0]{\@secondoftwo}%
\providecommand \translation [1]{[#1]}%
\providecommand \BibitemOpen [0]{}%
\providecommand \bibitemStop [0]{}%
\providecommand \bibitemNoStop [0]{.\EOS\space}%
\providecommand \EOS [0]{\spacefactor3000\relax}%
\providecommand \BibitemShut  [1]{\csname bibitem#1\endcsname}%
\let\auto@bib@innerbib\@empty
%</preamble>
\bibitem [{\citenamefont {Henn}\ \emph {et~al.}(2009)\citenamefont {Henn},
  \citenamefont {Seman}, \citenamefont {Roati}, \citenamefont {Magalh\~{a}es},\
  and\ \citenamefont {Bagnato}}]{Henn2009}%
  \BibitemOpen
  \bibfield  {author} {\bibinfo {author} {\bibfnamefont {E.}~\bibnamefont
  {Henn}}, \bibinfo {author} {\bibfnamefont {J.}~\bibnamefont {Seman}},
  \bibinfo {author} {\bibfnamefont {G.}~\bibnamefont {Roati}}, \bibinfo
  {author} {\bibfnamefont {K.}~\bibnamefont {Magalh\~{a}es}}, \ and\ \bibinfo
  {author} {\bibfnamefont {V.}~\bibnamefont {Bagnato}},\ }\href {\doibase
  10.1103/PhysRevLett.103.045301} {\bibfield  {journal} {\bibinfo  {journal}
  {Phys. Rev. Lett.}\ }\textbf {\bibinfo {volume} {103}},\ \bibinfo {pages} {1}
  (\bibinfo {year} {2009})}\BibitemShut {NoStop}%
\bibitem [{\citenamefont {Serafini}\ \emph {et~al.}(2015)\citenamefont
  {Serafini}, \citenamefont {Barbiero}, \citenamefont {Debortoli},
  \citenamefont {Donadello}, \citenamefont {Larcher}, \citenamefont {Dalfovo},
  \citenamefont {Lamporesi},\ and\ \citenamefont
  {Ferrari}}]{Serafini2015_VortexBEC}%
  \BibitemOpen
  \bibfield  {author} {\bibinfo {author} {\bibfnamefont {S.}~\bibnamefont
  {Serafini}}, \bibinfo {author} {\bibfnamefont {M.}~\bibnamefont {Barbiero}},
  \bibinfo {author} {\bibfnamefont {M.}~\bibnamefont {Debortoli}}, \bibinfo
  {author} {\bibfnamefont {S.}~\bibnamefont {Donadello}}, \bibinfo {author}
  {\bibfnamefont {F.}~\bibnamefont {Larcher}}, \bibinfo {author} {\bibfnamefont
  {F.}~\bibnamefont {Dalfovo}}, \bibinfo {author} {\bibfnamefont
  {G.}~\bibnamefont {Lamporesi}}, \ and\ \bibinfo {author} {\bibfnamefont
  {G.}~\bibnamefont {Ferrari}},\ }\href {\doibase
  10.1103/PhysRevLett.115.170402} {\bibfield  {journal} {\bibinfo  {journal}
  {Phys. Rev. Lett.}\ }\textbf {\bibinfo {volume} {115}},\ \bibinfo {pages}
  {170402} (\bibinfo {year} {2015})}\BibitemShut {NoStop}%
%
\bibitem [{\citenamefont {Bewley}\ \emph {et~al.}(2006)\citenamefont {Bewley},
  \citenamefont {Lathrop},\ and\ \citenamefont {Sreenivasan}}]{Bewley2006}%
  \BibitemOpen
  \bibfield  {author} {\bibinfo {author} {\bibfnamefont {G.~P.}\ \bibnamefont
  {Bewley}}, \bibinfo {author} {\bibfnamefont {D.~P.}\ \bibnamefont {Lathrop}},
  \ and\ \bibinfo {author} {\bibfnamefont {K.~R.}\ \bibnamefont
  {Sreenivasan}},\ }\href {\doibase 10.1038/441588a} {\bibfield  {journal}
  {\bibinfo  {journal} {Nature}\ }\textbf {\bibinfo {volume} {441}},\ \bibinfo
  {pages} {588} (\bibinfo {year} {2006})}\BibitemShut {NoStop}%
\bibitem [{\citenamefont {La~Mantia}\ and\ \citenamefont
  {Skrbek}(2014)}]{MantiaVisPart}%
  \BibitemOpen
  \bibfield  {author} {\bibinfo {author} {\bibfnamefont {M.}~\bibnamefont
  {La~Mantia}}\ and\ \bibinfo {author} {\bibfnamefont {L.}~\bibnamefont
  {Skrbek}},\ }\href {\doibase 10.1103/PhysRevB.90.014519} {\bibfield
  {journal} {\bibinfo  {journal} {Phys. Rev. B}\ }\textbf {\bibinfo {volume}
  {90}},\ \bibinfo {pages} {014519} (\bibinfo {year} {2014})}\BibitemShut
  {NoStop}%
\bibitem [{\citenamefont {La~Mantia}\ \emph {et~al.}(2013)\citenamefont
  {La~Mantia}, \citenamefont {Duda}, \citenamefont {Rotter},\ and\
  \citenamefont {Skrbek}}]{la2013lagrangian}%
  \BibitemOpen
  \bibfield  {author} {\bibinfo {author} {\bibfnamefont {M.}~\bibnamefont
  {La~Mantia}}, \bibinfo {author} {\bibfnamefont {D.}~\bibnamefont {Duda}},
  \bibinfo {author} {\bibfnamefont {M.}~\bibnamefont {Rotter}}, \ and\ \bibinfo
  {author} {\bibfnamefont {L.}~\bibnamefont {Skrbek}},\ }\href@noop {}
  {\bibfield  {journal} {\bibinfo  {journal} {J. of Fluid Mech.}\
  }\textbf {\bibinfo {volume} {717}},\ \bibinfo {pages} {R9} (\bibinfo {year}
  {2013})}\BibitemShut {NoStop}%
\bibitem [{\citenamefont {Gao}\ \emph {et~al.}(2015)\citenamefont {Gao},
  \citenamefont {Marakov}, \citenamefont {Guo}, \citenamefont {Pawlowski},
  \citenamefont {{Van Sciver}}, \citenamefont {Ihas}, \citenamefont
  {McKinsey},\ and\ \citenamefont {Vinen}}]{Gao2015}%
  \BibitemOpen
  \bibfield  {author} {\bibinfo {author} {\bibfnamefont {J.}~\bibnamefont
  {Gao}}, \bibinfo {author} {\bibfnamefont {a.}~\bibnamefont {Marakov}},
  \bibinfo {author} {\bibfnamefont {W.}~\bibnamefont {Guo}}, \bibinfo {author}
  {\bibfnamefont {B.~T.}\ \bibnamefont {Pawlowski}}, \bibinfo {author}
  {\bibfnamefont {S.~W.}\ \bibnamefont {{Van Sciver}}}, \bibinfo {author}
  {\bibfnamefont {G.~G.}\ \bibnamefont {Ihas}}, \bibinfo {author}
  {\bibfnamefont {D.~N.}\ \bibnamefont {McKinsey}}, \ and\ \bibinfo {author}
  {\bibfnamefont {W.~F.}\ \bibnamefont {Vinen}},\ }\href {\doibase
  10.1063/1.4930147} {\bibfield  {journal} {\bibinfo  {journal} {Review of
  Scientific Instruments}\ }\textbf {\bibinfo {volume} {86}},\ \bibinfo {pages}
  {093904} (\bibinfo {year} {2015})}\BibitemShut {NoStop}%
\bibitem [{\citenamefont {Zmeev}\ \emph {et~al.}(2013)\citenamefont {Zmeev},
  \citenamefont {Pakpour}, \citenamefont {Walmsley}, \citenamefont {Golov},
  \citenamefont {Guo}, \citenamefont {McKinsey}, \citenamefont {Ihas},
  \citenamefont {McClintock}, \citenamefont {Fisher},\ and\ \citenamefont
  {Vinen}}]{Zmeev2013}%
  \BibitemOpen
  \bibfield  {author} {\bibinfo {author} {\bibfnamefont {D.~E.}\ \bibnamefont
  {Zmeev}}, \bibinfo {author} {\bibfnamefont {F.}~\bibnamefont {Pakpour}},
  \bibinfo {author} {\bibfnamefont {P.~M.}\ \bibnamefont {Walmsley}}, \bibinfo
  {author} {\bibfnamefont {A.~I.}\ \bibnamefont {Golov}}, \bibinfo {author}
  {\bibfnamefont {W.}~\bibnamefont {Guo}}, \bibinfo {author} {\bibfnamefont
  {D.~N.}\ \bibnamefont {McKinsey}}, \bibinfo {author} {\bibfnamefont {G.~G.}\
  \bibnamefont {Ihas}}, \bibinfo {author} {\bibfnamefont {P.~V.~E.}\
  \bibnamefont {McClintock}}, \bibinfo {author} {\bibfnamefont {S.~N.}\
  \bibnamefont {Fisher}}, \ and\ \bibinfo {author} {\bibfnamefont {W.~F.}\
  \bibnamefont {Vinen}},\ }\href {\doibase 10.1103/PhysRevLett.110.175303}
  {\bibfield  {journal} {\bibinfo  {journal} {Phys. Rev. Lett.}\ }\textbf
  {\bibinfo {volume} {110}},\ \bibinfo {pages} {175303} (\bibinfo {year}
  {2013})}\BibitemShut {NoStop}%
\bibitem [{\citenamefont {Frisch}(1995)}]{Frisch1995}%
  \BibitemOpen
  \bibfield  {author} {\bibinfo {author} {\bibfnamefont {U.}~\bibnamefont
  {Frisch}},\ }\href@noop {} {\emph {\bibinfo {title} {{Turbulence: The Legacy
  of A. N. Kolmogorov}}}}\ (\bibinfo  {publisher} {Cambridge University
  Press},\ \bibinfo {year} {1995})\BibitemShut {NoStop}%
\bibitem [{\citenamefont {Maurer}\ and\ \citenamefont
  {Tabeling}(1998{\natexlab{a}})}]{maurer1998local}%
  \BibitemOpen
  \bibfield  {author} {\bibinfo {author} {\bibfnamefont {J.}~\bibnamefont
  {Maurer}}\ and\ \bibinfo {author} {\bibfnamefont {P.}~\bibnamefont
  {Tabeling}},\ }\href@noop {} {\bibfield  {journal} {\bibinfo  {journal} {EPL
  (Europhysics Letters)}\ }\textbf {\bibinfo {volume} {43}},\ \bibinfo {pages}
  {29} (\bibinfo {year} {1998}{\natexlab{a}})}\BibitemShut {NoStop}%
\bibitem [{\citenamefont {Salort}\ \emph {et~al.}(2010)\citenamefont {Salort},
  \citenamefont {Baudet}, \citenamefont {Castaing}, \citenamefont {Chabaud},
  \citenamefont {Daviaud}, \citenamefont {Didelot}, \citenamefont {Diribarne},
  \citenamefont {Dubrulle}, \citenamefont {Gagne}, \citenamefont {Gauthier}
  \emph {et~al.}}]{salort2010turbulent}%
  \BibitemOpen
  \bibfield  {author} {\bibinfo {author} {\bibfnamefont {J.}~\bibnamefont
  {Salort}}, \bibinfo {author} {\bibfnamefont {C.}~\bibnamefont {Baudet}},
  \bibinfo {author} {\bibfnamefont {B.}~\bibnamefont {Castaing}}, \bibinfo
  {author} {\bibfnamefont {B.}~\bibnamefont {Chabaud}}, \bibinfo {author}
  {\bibfnamefont {F.}~\bibnamefont {Daviaud}}, \bibinfo {author} {\bibfnamefont
  {T.}~\bibnamefont {Didelot}}, \bibinfo {author} {\bibfnamefont
  {P.}~\bibnamefont {Diribarne}}, \bibinfo {author} {\bibfnamefont
  {B.}~\bibnamefont {Dubrulle}}, \bibinfo {author} {\bibfnamefont
  {Y.}~\bibnamefont {Gagne}}, \bibinfo {author} {\bibfnamefont
  {F.}~\bibnamefont {Gauthier}},  \emph {et~al.},\ }\href@noop {} {\bibfield
  {journal} {\bibinfo  {journal} {Physi. of Fluids (1994-present)}\ }\textbf
  {\bibinfo {volume} {22}},\ \bibinfo {pages} {125102} (\bibinfo {year}
  {2010})}\BibitemShut {NoStop}%
\bibitem [{\citenamefont {Vinen}\ and\ \citenamefont
  {Niemela}(2002)}]{Vinen2002}%
  \BibitemOpen
  \bibfield  {author} {\bibinfo {author} {\bibfnamefont {W.~F.}\ \bibnamefont
  {Vinen}}\ and\ \bibinfo {author} {\bibfnamefont {J.~J.}\ \bibnamefont
  {Niemela}},\ }\href {\doibase 10.1023/A:1019695418590} {\bibfield  {journal}
  {\bibinfo  {journal} {J. Low Temp. Phys.}\ }\textbf {\bibinfo {volume}
  {128}},\ \bibinfo {pages} {167} (\bibinfo {year} {2002})}\BibitemShut
  {NoStop}%
\bibitem [{\citenamefont {Walmsley}\ \emph {et~al.}(2007)\citenamefont
  {Walmsley}, \citenamefont {Golov}, \citenamefont {Hall}, \citenamefont
  {Levchenko},\ and\ \citenamefont {Vinen}}]{Walmsley2007}%
  \BibitemOpen
  \bibfield  {author} {\bibinfo {author} {\bibfnamefont {P.~M.}\ \bibnamefont
  {Walmsley}}, \bibinfo {author} {\bibfnamefont {A.~I.}\ \bibnamefont {Golov}},
  \bibinfo {author} {\bibfnamefont {H.~E.}\ \bibnamefont {Hall}}, \bibinfo
  {author} {\bibfnamefont {A.~A.}\ \bibnamefont {Levchenko}}, \ and\ \bibinfo
  {author} {\bibfnamefont {W.~F.}\ \bibnamefont {Vinen}},\ }\href {\doibase
  10.1103/PhysRevLett.99.265302} {\bibfield  {journal} {\bibinfo  {journal}
  {Phys. Rev. Lett.}\ }\textbf {\bibinfo {volume} {99}},\ \bibinfo {pages}
  {265302} (\bibinfo {year} {2007})}\BibitemShut {NoStop}%
\bibitem [{\citenamefont {Nore}\ \emph
  {et~al.}(1997{\natexlab{a}})\citenamefont {Nore}, \citenamefont {Abid},\ and\
  \citenamefont {Brachet}}]{Nore1997a}%
  \BibitemOpen
  \bibfield  {author} {\bibinfo {author} {\bibfnamefont {C.}~\bibnamefont
  {Nore}}, \bibinfo {author} {\bibfnamefont {M.}~\bibnamefont {Abid}}, \ and\
  \bibinfo {author} {\bibfnamefont {M.E.}~\bibnamefont {Brachet}},\ }\href
  {\doibase 10.1103/PhysRevLett.78.3896} {\bibfield  {journal} {\bibinfo
  {journal} {Phys. Rev. Lett.}\ }\textbf {\bibinfo {volume} {78}},\ \bibinfo
  {pages} {3896} (\bibinfo {year} {1997}{\natexlab{a}})}\BibitemShut {NoStop}%
%
\bibitem{Footenote1} Using a Pad\'e approximation for the vortex profile (see [30]), the coherence length $\xi$ corresponds to a vortex core of size $a$ defined  as $\rho(r=a)=0.27\rho_\infty$, where $\rho_\infty$ is the value of the density far away from the vortex.
%
\bibitem [{\citenamefont {Yepez}\ \emph {et~al.}(2009)\citenamefont {Yepez},
  \citenamefont {Vahala}, \citenamefont {Vahala},\ and\ \citenamefont
  {Soe}}]{Yepez2009}%
  \BibitemOpen
  \bibfield  {author} {\bibinfo {author} {\bibfnamefont {J.}~\bibnamefont
  {Yepez}}, \bibinfo {author} {\bibfnamefont {G.}~\bibnamefont {Vahala}},
  \bibinfo {author} {\bibfnamefont {L.}~\bibnamefont {Vahala}}, \ and\ \bibinfo
  {author} {\bibfnamefont {M.}~\bibnamefont {Soe}},\ }\href {\doibase
  10.1103/PhysRevLett.103.084501} {\bibfield  {journal} {\bibinfo  {journal}
  {Phys. Rev. Lett.}\ }\textbf {\bibinfo {volume} {103}},\ \bibinfo {pages} {3}
  (\bibinfo {year} {2009})}\BibitemShut {NoStop}%
\bibitem [{\citenamefont {Salort}\ \emph {et~al.}(2012)\citenamefont {Salort},
  \citenamefont {Chabaud}, \citenamefont {L\'{e}v\^{e}que},\ and\ \citenamefont
  {Roche}}]{Salort2012}%
  \BibitemOpen
  \bibfield  {author} {\bibinfo {author} {\bibfnamefont {J.}~\bibnamefont
  {Salort}}, \bibinfo {author} {\bibfnamefont {B.}~\bibnamefont {Chabaud}},
  \bibinfo {author} {\bibfnamefont {E.}~\bibnamefont {L\'{e}v\^{e}que}}, \ and\
  \bibinfo {author} {\bibfnamefont {P.-E.}\ \bibnamefont {Roche}},\ }\href
  {\doibase 10.1209/0295-5075/97/34006} {\bibfield  {journal} {\bibinfo
  {journal} {EPL (Europhysics Letters)}\ }\textbf {\bibinfo {volume} {97}},\
  \bibinfo {pages} {34006} (\bibinfo {year} {2012})}\BibitemShut {NoStop}%
%
\bibitem [{\citenamefont {Baggaley}\ and\ \citenamefont
  {Barenghi}(2011)}]{Baggaley2011a}%
  \BibitemOpen
  \bibfield  {author} {\bibinfo {author} {\bibfnamefont {A.W}~\bibnamefont
  {Baggaley}}\ and\ \bibinfo {author} {\bibfnamefont {C.F}~\bibnamefont
  {Barenghi}},\ }\href {\doibase 10.1103/PhysRevE.84.067301} {\bibfield
  {journal} {\bibinfo  {journal} {Phys. Rev. E}\ }\textbf {\bibinfo {volume}
  {84}},\ \bibinfo {pages} {067301} (\bibinfo {year} {2011})}\BibitemShut {NoStop}%
%
\bibitem [{\citenamefont {Comte-Bellot}\ and\ \citenamefont
  {Corrsin}(1966)}]{comte1966use}%
  \BibitemOpen
  \bibfield  {author} {\bibinfo {author} {\bibfnamefont {G.}~\bibnamefont
  {Comte-Bellot}}\ and\ \bibinfo {author} {\bibfnamefont {S.}~\bibnamefont
  {Corrsin}},\ }\href@noop {} {\bibfield  {journal} {\bibinfo  {journal}
  {Journal of Fluid Mechanics}\ }\textbf {\bibinfo {volume} {25}},\ \bibinfo
  {pages} {657} (\bibinfo {year} {1966})}\BibitemShut {NoStop}%
\bibitem [{\citenamefont {Pope}(2000)}]{pope2000turbulent}%
  \BibitemOpen
  \bibfield  {author} {\bibinfo {author} {\bibfnamefont {S.~B.}\ \bibnamefont
  {Pope}},\ }\href@noop {} {\emph {\bibinfo {title} {Turbulent flows}}}\
  (\bibinfo  {publisher} {Cambridge university press},\ \bibinfo {year}
  {2000})\BibitemShut {NoStop}%
\bibitem [{\citenamefont {Stalp}\ \emph {et~al.}(1999)\citenamefont {Stalp},
  \citenamefont {Skrbek},\ and\ \citenamefont {Donnelly}}]{stalp1999decay}%
  \BibitemOpen
  \bibfield  {author} {\bibinfo {author} {\bibfnamefont {S.~R.}\ \bibnamefont
  {Stalp}}, \bibinfo {author} {\bibfnamefont {L.}~\bibnamefont {Skrbek}}, \
  and\ \bibinfo {author} {\bibfnamefont {R.~J.}\ \bibnamefont {Donnelly}},\
  }\href@noop {} {\bibfield  {journal} {\bibinfo  {journal} {Physical review
  letters}\ }\textbf {\bibinfo {volume} {82}},\ \bibinfo {pages} {4831}
  (\bibinfo {year} {1999})}\BibitemShut {NoStop}%
\bibitem [{\citenamefont {Zmeev}\ \emph {et~al.}(2015)\citenamefont {Zmeev},
  \citenamefont {Walmsley}, \citenamefont {Golov}, \citenamefont {McClintock},
  \citenamefont {Fisher},\ and\ \citenamefont {Vinen}}]{Zmeev2015}%
  \BibitemOpen
  \bibfield  {author} {\bibinfo {author} {\bibfnamefont {D.~E.}\ \bibnamefont
  {Zmeev}}, \bibinfo {author} {\bibfnamefont {P.~M.}\ \bibnamefont {Walmsley}},
  \bibinfo {author} {\bibfnamefont {A.~I.}\ \bibnamefont {Golov}}, \bibinfo
  {author} {\bibfnamefont {P.~V.~E.}\ \bibnamefont {McClintock}}, \bibinfo
  {author} {\bibfnamefont {S.~N.}\ \bibnamefont {Fisher}}, \ and\ \bibinfo
  {author} {\bibfnamefont {W.~F.}\ \bibnamefont {Vinen}},\ }\href {\doibase
  10.1103/PhysRevLett.115.155303} {\bibfield  {journal} {\bibinfo  {journal}
  {Phys. Rev. Lett.}\ }\textbf {\bibinfo {volume} {115}},\ \bibinfo {pages}
  {155303} (\bibinfo {year} {2015})}\BibitemShut {NoStop}%
\bibitem [{\citenamefont {Bradley}\ \emph {et~al.}(2012)\citenamefont
  {Bradley}, \citenamefont {Fisher}, \citenamefont {Gu\'enault}, \citenamefont
  {Haley}, \citenamefont {Kumar}, \citenamefont {Lawson}, \citenamefont
  {Schanen}, \citenamefont {McClintock}, \citenamefont {Munday}, \citenamefont
  {Pickett}, \citenamefont {Poole}, \citenamefont {Tsepelin},\ and\
  \citenamefont {Williams}}]{Bradley2012}%
  \BibitemOpen
  \bibfield  {author} {\bibinfo {author} {\bibfnamefont {D.~I.}\ \bibnamefont
  {Bradley}}, \bibinfo {author} {\bibfnamefont {S.~N.}\ \bibnamefont {Fisher}},
  \bibinfo {author} {\bibfnamefont {A.~M.}\ \bibnamefont {Gu\'enault}},
  \bibinfo {author} {\bibfnamefont {R.~P.}\ \bibnamefont {Haley}}, \bibinfo
  {author} {\bibfnamefont {M.}~\bibnamefont {Kumar}}, \bibinfo {author}
  {\bibfnamefont {C.~R.}\ \bibnamefont {Lawson}}, \bibinfo {author}
  {\bibfnamefont {R.}~\bibnamefont {Schanen}}, \bibinfo {author} {\bibfnamefont
  {P.~V.~E.}\ \bibnamefont {McClintock}}, \bibinfo {author} {\bibfnamefont
  {L.}~\bibnamefont {Munday}}, \bibinfo {author} {\bibfnamefont {G.~R.}\
  \bibnamefont {Pickett}}, \bibinfo {author} {\bibfnamefont {M.}~\bibnamefont
  {Poole}}, \bibinfo {author} {\bibfnamefont {V.}~\bibnamefont {Tsepelin}}, \
  and\ \bibinfo {author} {\bibfnamefont {P.}~\bibnamefont {Williams}},\ }\href
  {\doibase 10.1103/PhysRevB.85.224533} {\bibfield  {journal} {\bibinfo
  {journal} {Phys. Rev. B}\ }\textbf {\bibinfo {volume} {85}},\ \bibinfo
  {pages} {224533} (\bibinfo {year} {2012})}\BibitemShut {NoStop}%
\bibitem [{\citenamefont {Bradley}\ \emph {et~al.}(2005)\citenamefont
  {Bradley}, \citenamefont {Clubb}, \citenamefont {Fisher}, \citenamefont
  {Gu\'enault}, \citenamefont {Haley}, \citenamefont {Matthews}, \citenamefont
  {Pickett}, \citenamefont {Tsepelin},\ and\ \citenamefont
  {Zaki}}]{Bradley2005}%
  \BibitemOpen
  \bibfield  {author} {\bibinfo {author} {\bibfnamefont {D.~I.}\ \bibnamefont
  {Bradley}}, \bibinfo {author} {\bibfnamefont {D.~O.}\ \bibnamefont {Clubb}},
  \bibinfo {author} {\bibfnamefont {S.~N.}\ \bibnamefont {Fisher}}, \bibinfo
  {author} {\bibfnamefont {A.~M.}\ \bibnamefont {Gu\'enault}}, \bibinfo
  {author} {\bibfnamefont {R.~P.}\ \bibnamefont {Haley}}, \bibinfo {author}
  {\bibfnamefont {C.~J.}\ \bibnamefont {Matthews}}, \bibinfo {author}
  {\bibfnamefont {G.~R.}\ \bibnamefont {Pickett}}, \bibinfo {author}
  {\bibfnamefont {V.}~\bibnamefont {Tsepelin}}, \ and\ \bibinfo {author}
  {\bibfnamefont {K.}~\bibnamefont {Zaki}},\ }\href {\doibase
  10.1103/PhysRevLett.95.035302} {\bibfield  {journal} {\bibinfo  {journal}
  {Phys. Rev. Lett.}\ }\textbf {\bibinfo {volume} {95}},\ \bibinfo {pages}
  {035302} (\bibinfo {year} {2005})}\BibitemShut {NoStop}%
\bibitem [{\citenamefont {Nore}\ \emph
  {et~al.}(1997{\natexlab{b}})\citenamefont {Nore}, \citenamefont {Abid},\ and\
  \citenamefont {Brachet}}]{Nore1997}%
  \BibitemOpen
  \bibfield  {author} {\bibinfo {author} {\bibfnamefont {C.}~\bibnamefont
  {Nore}}, \bibinfo {author} {\bibfnamefont {M.}~\bibnamefont {Abid}}, \ and\
  \bibinfo {author} {\bibfnamefont {M.~E.}\ \bibnamefont {Brachet}},\ }\href
  {\doibase 10.1063/1.869473} {\bibfield  {journal} {\bibinfo  {journal}
  {Physics of Fluids}\ }\textbf {\bibinfo {volume} {9}},\ \bibinfo {pages}
  {2644} (\bibinfo {year} {1997}{\natexlab{b}})}\BibitemShut {NoStop}%
\bibitem [{\citenamefont {Reeves}\ \emph {et~al.}(2013)\citenamefont {Reeves},
  \citenamefont {Billam}, \citenamefont {Anderson},\ and\ \citenamefont
  {Bradley}}]{Reeves2013}%
  \BibitemOpen
  \bibfield  {author} {\bibinfo {author} {\bibfnamefont {M.T.}~\bibnamefont
  {Reeves}}, \bibinfo {author} {\bibfnamefont {T.P.}~\bibnamefont {Billam}},
  \bibinfo {author} {\bibfnamefont {B.P.}~\bibnamefont {Anderson}}, \ and\
  \bibinfo {author} {\bibfnamefont {A.S.}~\bibnamefont {Bradley}},\ }\href
  {\doibase 10.1103/PhysRevLett.110.104501} {\bibfield  {journal} {\bibinfo
  {journal} {Phys. Rev. Lett.}\ }\textbf {\bibinfo {volume} {110}},\ \bibinfo
  {pages} {104501} (\bibinfo {year} {2013})}\BibitemShut {NoStop}%
\bibitem [{\citenamefont {Krstulovic}\ and\ \citenamefont
  {Brachet}(2011{\natexlab{b}})}]{Krstulovic2011a}%
  \BibitemOpen
  \bibfield  {author} {\bibinfo {author} {\bibfnamefont {G.}~\bibnamefont
  {Krstulovic}}\ and\ \bibinfo {author} {\bibfnamefont {M.}~\bibnamefont
  {Brachet}},\ }\href {\doibase 10.1103/PhysRevLett.106.115303} {\bibfield
  {journal} {\bibinfo  {journal} {Phys. Rev. Lett.}\ }\textbf {\bibinfo
  {volume} {106}},\ \bibinfo {pages} {2} (\bibinfo {year}
  {2011}{\natexlab{b}})}\BibitemShut {NoStop}%
\bibitem [{\citenamefont {Shukla}\ \emph {et~al.}(2013)\citenamefont {Shukla},
  \citenamefont {Brachet},\ and\ \citenamefont {Pandit}}]{Shukla2013}%
  \BibitemOpen
  \bibfield  {author} {\bibinfo {author} {\bibfnamefont {V.}~\bibnamefont
  {Shukla}}, \bibinfo {author} {\bibfnamefont {M.}~\bibnamefont {Brachet}}, \
  and\ \bibinfo {author} {\bibfnamefont {R.}~\bibnamefont {Pandit}},\ }\href
  {\doibase 10.1088/1367-2630/15/11/113025} {\bibfield  {journal} {\bibinfo
  {journal} {New J. of Phys.}\ }\textbf {\bibinfo {volume} {15}} (\bibinfo
  {year} {2013}),\ 10.1088/1367-2630/15/11/113025},\ \Eprint
  {http://arxiv.org/abs/1301.3383} {1301.3383} \BibitemShut {NoStop}%
\bibitem [{\citenamefont {Huepe}\ and\ \citenamefont
  {Brachet}(2000)}]{huepe2000nucleation}%
  \BibitemOpen
  \bibfield  {author} {\bibinfo {author} {\bibfnamefont {C.}~\bibnamefont
  {Huepe}}\ and\ \bibinfo {author} {\bibfnamefont {M.-E.}\ \bibnamefont
  {Brachet}},\ }\href@noop {} {\bibfield  {journal} {\bibinfo  {journal}
  {Physica D: Nonlinear Phenomena}\ }\textbf {\bibinfo {volume} {140}},\
  \bibinfo {pages} {126} (\bibinfo {year} {2000})}\BibitemShut {NoStop}%
\bibitem [{\citenamefont {Frisch}\ \emph {et~al.}(1992)\citenamefont {Frisch},
  \citenamefont {Pomeau},\ and\ \citenamefont {Rica}}]{frisch1992transition}%
  \BibitemOpen
  \bibfield  {author} {\bibinfo {author} {\bibfnamefont {T.}~\bibnamefont
  {Frisch}}, \bibinfo {author} {\bibfnamefont {Y.}~\bibnamefont {Pomeau}}, \
  and\ \bibinfo {author} {\bibfnamefont {S.}~\bibnamefont {Rica}},\ }\href@noop
  {} {\bibfield  {journal} {\bibinfo  {journal} {Physical review letters}\
  }\textbf {\bibinfo {volume} {69}},\ \bibinfo {pages} {1644} (\bibinfo {year}
  {1992})}\BibitemShut {NoStop}%
  \bibitem [{\citenamefont {Pismen}(1999)}]{pismen1999vortices}%
  \BibitemOpen
  \bibfield  {author} {\bibinfo {author} {\bibfnamefont {L.~M.}\ \bibnamefont
  {Pismen}},\ }\href@noop {} {\emph {\bibinfo {title} {Vortices in nonlinear
  fields: From liquid crystals to superfluids, from non-equilibrium patterns to
  cosmic strings}}},\ Vol.\ \bibinfo {volume} {100}\ (\bibinfo  {publisher}
  {Oxford University Press},\ \bibinfo {year} {1999})\BibitemShut {NoStop}%
\bibitem [{\citenamefont {Sasaki}\ \emph {et~al.}(2010)\citenamefont {Sasaki},
  \citenamefont {Suzuki},\ and\ \citenamefont {Saito}}]{SasakiVonKarmanGP2d}%
  \BibitemOpen
  \bibfield  {author} {\bibinfo {author} {\bibfnamefont {K.}~\bibnamefont
  {Sasaki}}, \bibinfo {author} {\bibfnamefont {N.}~\bibnamefont {Suzuki}}, \
  and\ \bibinfo {author} {\bibfnamefont {H.}~\bibnamefont {Saito}},\ }\href
  {\doibase 10.1103/PhysRevLett.104.150404} {\bibfield  {journal} {\bibinfo
  {journal} {Phys. Rev. Lett.}\ }\textbf {\bibinfo {volume} {104}},\ \bibinfo
  {pages} {150404} (\bibinfo {year} {2010})}\BibitemShut {NoStop}%
\bibitem [{\citenamefont {Krstulovic}\ and\ \citenamefont
  {Brachet}(2011{\natexlab{a}})}]{Krstulovic2011b}%
  \BibitemOpen
  \bibfield  {author} {\bibinfo {author} {\bibfnamefont {G.}~\bibnamefont
  {Krstulovic}}\ and\ \bibinfo {author} {\bibfnamefont {M.E.}~\bibnamefont
  {Brachet}},\ }\href@noop {} {\bibfield  {journal} {\bibinfo  {journal}
  {Phys. Rev. E}\ }\textbf {\bibinfo {volume} {83}},\ \bibinfo {pages}
  {066311} (\bibinfo {year} {2011}{\natexlab{a}})}\BibitemShut {NoStop}%
\bibitem [{\citenamefont {Vinen}(1957)}]{Vinen493}%
  \BibitemOpen
  \bibfield  {author} {\bibinfo {author} {\bibfnamefont {W.~F.}\ \bibnamefont
  {Vinen}},\ }\href {\doibase 10.1098/rspa.1957.0191} {\bibfield  {journal}
  {\bibinfo  {journal} {Proceedings of the Royal Society of London A:
  Mathematical, Physical and Engineering Sciences}\ }\textbf {\bibinfo {volume}
  {242}},\ \bibinfo {pages} {493} (\bibinfo {year} {1957})}\BibitemShut
  {NoStop}%
\bibitem [{\citenamefont {{La Mantia}}\ and\ \citenamefont
  {Skrbek}(2014)}]{LaMantia2014}%
  \BibitemOpen
  \bibfield  {author} {\bibinfo {author} {\bibfnamefont {M.}~\bibnamefont {{La
  Mantia}}}\ and\ \bibinfo {author} {\bibfnamefont {L.}~\bibnamefont
  {Skrbek}},\ }\href {\doibase 10.1209/0295-5075/105/46002} {\bibfield
  {journal} {\bibinfo  {journal} {Europhys. Lett.}\ }\textbf {\bibinfo {volume}
  {105}},\ \bibinfo {pages} {46002} (\bibinfo {year} {2014})}\BibitemShut
  {NoStop}%
\bibitem [{\citenamefont {White}\ \emph {et~al.}(2010)\citenamefont {White},
  \citenamefont {Barenghi}, \citenamefont {Proukakis}, \citenamefont {Youd},\
  and\ \citenamefont {Wacks}}]{White2010}%
  \BibitemOpen
  \bibfield  {author} {\bibinfo {author} {\bibfnamefont {A.~C.}\ \bibnamefont
  {White}}, \bibinfo {author} {\bibfnamefont {C.~F.}\ \bibnamefont {Barenghi}},
  \bibinfo {author} {\bibfnamefont {N.~P.}\ \bibnamefont {Proukakis}}, \bibinfo
  {author} {\bibfnamefont {A.~J.}\ \bibnamefont {Youd}}, \ and\ \bibinfo
  {author} {\bibfnamefont {D.~H.}\ \bibnamefont {Wacks}},\ }\href {\doibase
  10.1103/PhysRevLett.104.075301} {\bibfield  {journal} {\bibinfo  {journal}
  {Phys. Rev. Lett.}\ }\textbf {\bibinfo {volume} {104}},\ \bibinfo {pages}
  {075301} (\bibinfo {year} {2010})}\BibitemShut {NoStop}%
\bibitem [{\citenamefont {She}\ and\ \citenamefont
  {Leveque}(1994)}]{she1994universal}%
  \BibitemOpen
  \bibfield  {author} {\bibinfo {author} {\bibfnamefont {Z.-S.}\ \bibnamefont
  {She}}\ and\ \bibinfo {author} {\bibfnamefont {E.}~\bibnamefont {Leveque}},\
  }\href@noop {} {\bibfield  {journal} {\bibinfo  {journal} {Physical review
  letters}\ }\textbf {\bibinfo {volume} {72}},\ \bibinfo {pages} {336}
  (\bibinfo {year} {1994})}\BibitemShut {NoStop}%
\bibitem [{\citenamefont {Salort}\ \emph {et~al.}(2011)\citenamefont {Salort},
  \citenamefont {Chabaud}, \citenamefont {L{\'e}v{\^e}que},\ and\ \citenamefont
  {Roche}}]{salort2011investigation}%
  \BibitemOpen
  \bibfield  {author} {\bibinfo {author} {\bibfnamefont {J.}~\bibnamefont
  {Salort}}, \bibinfo {author} {\bibfnamefont {B.}~\bibnamefont {Chabaud}},
  \bibinfo {author} {\bibfnamefont {E.}~\bibnamefont {L{\'e}v{\^e}que}}, \ and\
  \bibinfo {author} {\bibfnamefont {P.-E.}\ \bibnamefont {Roche}},\ }\bibfield
  {booktitle} {\emph {\bibinfo {booktitle} {Journal of Physics: Conference
  Series}},\ }\href@noop {} {\ \textbf {\bibinfo {volume} {318}},\ \bibinfo
  {pages} {042014} (\bibinfo {year} {2011})}\BibitemShut {NoStop}%
\bibitem [{\citenamefont {Bou{\'e}}\ \emph {et~al.}(2013)\citenamefont
  {Bou{\'e}}, \citenamefont {L'vov}, \citenamefont {Pomyalov},\ and\
  \citenamefont {Procaccia}}]{boue2013enhancement}%
  \BibitemOpen
  \bibfield  {author} {\bibinfo {author} {\bibfnamefont {L.}~\bibnamefont
  {Bou{\'e}}}, \bibinfo {author} {\bibfnamefont {V.}~\bibnamefont {L'vov}},
  \bibinfo {author} {\bibfnamefont {A.}~\bibnamefont {Pomyalov}}, \ and\
  \bibinfo {author} {\bibfnamefont {I.}~\bibnamefont {Procaccia}},\ }\href@noop
  {} {\bibfield  {journal} {\bibinfo  {journal} {Physical review letters}\
  }\textbf {\bibinfo {volume} {110}},\ \bibinfo {pages} {014502} (\bibinfo
  {year} {2013})}\BibitemShut {NoStop}%
\bibitem [{\citenamefont {Shukla}\ and\ \citenamefont
  {Pandit}(2015)}]{shukla2015multiscaling}%
  \BibitemOpen
  \bibfield  {author} {\bibinfo {author} {\bibfnamefont {V.}~\bibnamefont
  {Shukla}}\ and\ \bibinfo {author} {\bibfnamefont {R.}~\bibnamefont
  {Pandit}},\ }\href@noop {} {\bibfield  {journal} {\bibinfo  {journal} {arXiv
  preprint arXiv:1508.00448}\ } (\bibinfo {year} {2015})}\BibitemShut {NoStop}%
 %
 \bibitem [{\citenamefont {Lamporesi}\ \emph {et~al.}(2013)\citenamefont
  {Lamporesi}, \citenamefont {Donadello}, \citenamefont {Serafini},
  \citenamefont {Dalfovo},\ and\ \citenamefont
  {Ferrari}}]{Lamporesi2013VortexCreation}%
  \BibitemOpen
  \bibfield  {author} {\bibinfo {author} {\bibfnamefont {G.}~\bibnamefont
  {Lamporesi}}, \bibinfo {author} {\bibfnamefont {S.}~\bibnamefont
  {Donadello}}, \bibinfo {author} {\bibfnamefont {S.}~\bibnamefont {Serafini}},
  \bibinfo {author} {\bibfnamefont {F.}~\bibnamefont {Dalfovo}}, \ and\
  \bibinfo {author} {\bibfnamefont {G.}~\bibnamefont {Ferrari}},\ }\href
  {http://dx.doi.org/10.1038/nphys2734} {\bibfield  {journal} {\bibinfo
  {journal} {Nat Phys}\ }\textbf {\bibinfo {volume} {9}},\ \bibinfo {pages}
  {656} (\bibinfo {year} {2013})}\BibitemShut {NoStop}%
  \bibitem [{\citenamefont {Gottlieb}\ and\ \citenamefont
  {Orszag}(1977)}]{gottlieb1977numerical}%
  \BibitemOpen
  \bibfield  {author} {\bibinfo {author} {\bibfnamefont {D.}~\bibnamefont
  {Gottlieb}}\ and\ \bibinfo {author} {\bibfnamefont {S.~A.}\ \bibnamefont
  {Orszag}},\ }\href@noop {} {\emph {\bibinfo {title} {Numerical analysis of
  spectral methods: theory and applications}}},\ Vol.~\bibinfo {volume} {26}\
  (\bibinfo  {publisher} {Siam},\ \bibinfo {year} {1977})\BibitemShut {NoStop}%
\end{thebibliography}

\end{document}